\documentstyle[ltwol,latexsym,epsfig]{article}

\def\PRL{{\em Phys. Rev. Lett.} }
\def\PRB{{\em Phys. Rev.} B }

\begin{document}

\title{MAGNETIC FINGERPRINTS IN SUPERCONDUCTING Au$_{0.7}$In$_{0.3}$ CYLINDERS}

\author{Yu. ZADOROZHNY, D. R. HERMAN, AND Y. LIU}

\address{Department of Physics, The Pennsylvania State University, University Park, PA 16802, USA}

\twocolumn[\maketitle\abstracts{Reproducible, sample-specific magnetoresistance fluctuations (magnetic fingerprints) were observed in the low-temperature part of the superconducting transition regime of hollow Au$_{0.7}$In$_{0.3}$ cylinders of submicron diameter.  The amplitude of the fluctuations was found to exceed that of the universal conductance fluctuation in normal metals by several orders of magnitude.  The physical origin of these observations is related to mesoscopic fluctuations of the superconducting order parameter.}]

In the past two decades fascinating phenomena in normal-metal mesoscopic systems have been found and, for the most part, understood.\cite{1}  One of the most important aspects of mesoscopic physics is quantum interference over a length much larger than the atomic size.  In disordered mesoscopic samples, this remarkable phenomenon is manifested in seemingly random but fully reproducible, sample-specific magnetoresistance  fluctuations, referred to in the literature as magnetic fingerprints (MFPs).\cite{1}  These MFPs, which have emerged as a hallmark of mesoscopic physics, result from Aharonov-Bohm interference of electron waves.  Remarkably, the amplitude of the conductance fluctuations has a universal value of the order of $e^2/h$, known as the universal conductance fluctuation (UCF).  The physical origin of the UCF lies in the energy level statistics in disordered metal, in which the fluctuation in the number of energy levels within any energy interval is universally of the order of unity.\cite{1}

In the past few years, the UCF has also been examined in normal-metal samples in contact with one or more superconducting islands.\cite{2}  The superconducting pair potential penetrates inside the normal metal at a length characterized by the normal coherence length $L_T$ (proximity effect).  However, Andreev reflection from the normal metal-superconductor interfaces introduces electron-hole correlations in the normal metal that extend beyond $L_T$ and as far as the dephasing length $L_\phi$.\cite{3}  The additional phase coherence affects the conductance of the normal metal, leading to new physical phenomena.\cite{2}  However, no significant change was found in the amplitude of conductance fluctuations,\cite{4} as anticipated theoretically.\cite{5}

Interesting questions arise if superconductivity is introduced in the bulk, rather than at the boundary of a normal sample.  Consider a weakly disordered mesoscopic sample in which electrons become phase coherent well above the onset of superconductivity.  These phase-coherent normal electrons are extremely sensitive to impurity scattering.\cite{1}  However, when electrons form Cooper pairs, they become completely insensitive to randomness.  How do electrons respond to these opposite tendencies of motion?  In addition, in disordered metallic samples, energy levels fluctuate, leading to MFPs and the UCF as mentioned above.  What would the manifestation of the energy level fluctuation be in disordered superconductors?  We have carried out measurements on disordered superconducting Au-In cylinders to address these issues.

Au-In alloy has a rich phase diagram that includes compounds, AuIn and AuIn$_2$, and solid solutions with varying composition ratios.\cite{alloy}  In the bulk form, the maximum solid solubility of In in Au is only about 10\%.\cite{alloy}  When the In concentration exceeds this limit, a phase separation occurs, with the excess In forming In-rich grains.\cite{films}  The inhomogeneity in In concentration leads to spatially varying local $T_c$'s, resulting in a random superconductor-normal metal-superconductor Josephson junction array.  In the present study we concentrate on Au-In films with a nominal In fraction of 30\%.  Details of the sample preparation have been described elsewhere.\cite{finger}

\begin{figure}
\centerline{\epsfig{file=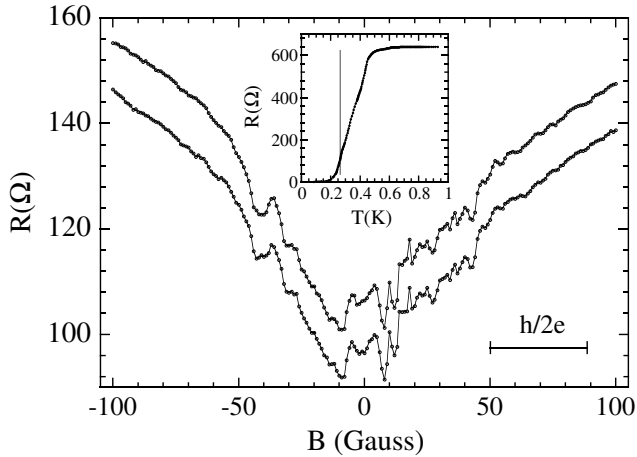,angle=0,width=3.4in}}
\caption{Two MR traces at $T = 0.25$~K, upper trace offset by $10~\Omega$ for clarity.  Inset: $R(T)$ in zero field.  The vertical line indicates the threshold temperature $T^{\ast}$ below which the MR fluctuation was found.}
\label{1}
\end{figure}

In Fig.~\ref{1} we show two traces of axial magnetoresistance (MR) taken back-to-back deep in the superconducting transition regime of a Au$_{0.7}$In$_{0.3}$ cylinder.  The diameter, film thickness, and the normal-state sheet resistance of this sample (Cylinder~12) were respectively 840~nm, 35~nm, and $1.7~\Omega$.  A non-periodic, asymmetric (with respect to the reversal of the magnetic field) MR pattern was found in both traces.  A comparison of the two traces shows a remarkable reproducibility of the pattern (the cross-correlation is 97\%).  This pattern can be seen as a reproducible resistance fluctuation, or a magnetic fingerprint, in a positive, symmetric MR background expected for a superconductor.  Similar MFPs have been found in other Au$_{0.7}$In$_{0.3}$ cylinders.

\begin{figure}
\centerline{\epsfig{file=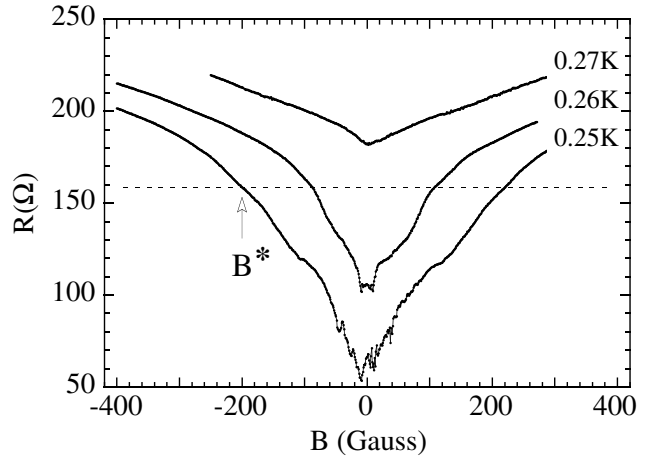,angle=0,width=3.4in}}
\caption{MR traces taken at temperatures as indicated.  MR fluctuation persists up to $B^{\ast}$.}
\label{2}
\end{figure}

A small increase in temperature was found to suppress the magnetoresistance fluctuation surprisingly strongly.  At temperature $T^{\ast} \approx 0.27$~K, indicated in the inset of Fig.~\ref{1} with a vertical line, the resistance fluctuation had already disappeared completely.  This trend is illustrated in Fig.~\ref{2}, in which we show MR traces taken at three different temperatures.  Magnetic field was found to have a similar effect.  Above a threshold field $B^{\ast}(T)$, the resistance fluctuation disappeared and the MR recovered the monotonic, symmetric behavior.  It is interesting to note that the fluctuation disappeared once the resistance was above a certain value, by increasing either temperature or magnetic field.

\begin{figure}
\centerline{\epsfig{file=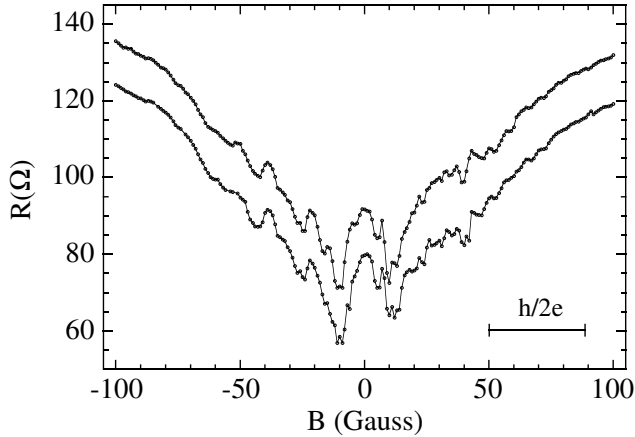,angle=0,width=3.4in}}
\caption{Two MR traces at $T = 0.25$~K, obtained after thermal cycling and featuring a different fluctuation pattern.  The upper trace is offset by $10~\Omega$.}
\label{3}
\end{figure}

The MFPs remained essentially the same in several consecutive scans.  However, after the sample was thermally cycled to around 10-15~K, well above the onset of superconductivity, a different fluctuation pattern was found, as shown for example in Fig.~\ref{2} (bottom trace) and in Fig.~\ref{3}.  The new MFP was again reproducible as illustrated by the two MR traces in Fig.~\ref{3}.  The magnitude of the zero-field resistance $R_{B=0}$ at fixed temperatures below $T^{\ast}$ was also found to change randomly as a result of thermal cycling.  The range of the resistance variation at $T=0.25$~K was about 10\% of the normal-state resistance $R_N$.  Applying a high (several Tesla) magnetic field also irreversibly changed the MR, similar to thermal cycling.

A resistance maximum at zero magnetic field is clearly seen in Fig.~\ref{3}.  Typically, superconducting fluctuations are suppressed by an  applied field, leading to a positive MR.  In the conventional Little-Parks (L-P) experiment, the MR at $B=0$ is always a minimum.\cite{L-P}  A negative MR as large as 25\% of $R_{B=0}$ deep in the superconducting transition regime is therefore very unusual.  Similar negative MR has been observed in other Au$_{0.7}$In$_{0.3}$ cylinders.  The negative MR was suppressed by a small temperature increase as all other features in the MR were (Fig.~\ref{2}).

The data shown in Figs.~\ref{1}-\ref{3} suggest that the conventional $h/2e$ L-P resistance oscillation was too weak to be observed or even absent.  Significant or complete suppression of resistance oscillations was found in all free-standing cylinders and can be attributed to spatial fluctuations in local $T_c$ along the sample circumference.\cite{finger}

Sample-specific MR could in principle result from multiple magnetic field driven transitions if the sample consisted of a collection of superconducting weak links with varying local critical field.  In this picture, however, successive suppression of superconductivity of each individual weak link as the (parallel) field increases would result in monotonic, step-like features in MR, accompanied by hysteresis.\cite{Adams}  Instead, MR of our samples was found to be strongly non-monotonic and non-hysteretic.  Furthermore, the MR was asymmetric with respect to the magnetic field reversal, which also can not be explained by the weak link picture.  All these considerations seem to suggest that superconducting weak links, if present in our samples, do not contribute significantly to the observed sample-specific MR.

Mesoscopic conductance fluctuations in {\it normal metals} are sensitive to impurity configurations, magnetic fields, and gate voltages.\cite{1}  Thermal cycling to moderately high temperature can affect the impurity configuration and therefore result in irreversible conductance change of the order of $e^2/h$.  Magnetic field of the order of the correlation field $B_{cor}$, corresponding to one flux quantum through the cross-section of the film, is required to change the conductance by $e^2/h$.\cite{1}  MFPs were also found in our samples, however, due to the suppression of superconductivity, the MFPs were only observed in fields up to $B^{\ast}$, smaller than $B_{cor} \approx 450$~G.  As a result, the most prominent fluctuation features had field scale much smaller than $B_{cor}$.  It should be noted that conductance fluctuations on field scales much smaller than $B_{cor}$ have been observed in normal-metal samples,\cite{6} with amplitude somewhat smaller than $e^2/h$.

The similarities between the sample-specific conductance in our samples and in mesoscopic normal-metal systems strongly suggest that the observed features are mesoscopic in origin.  However, the amplitude of these sample-specific conductance fluctuations appears to be much larger than that observed in normal samples.  An order-of-magnitude estimate gives $\Delta G = \Delta R_\Box / R_\Box^2 \approx 10^4 \, e^2/h$, where $R_\Box$ is the sheet resistance of the sample.

Theoretically, significantly enhanced sample-specific conductance fluctuations have been predicted for homogeneously disordered superconductors in the transition regime.  It has been shown that under appropriate conditions, such as close to the superconductor-insulator transition or in a strong parallel magnetic field, fluctuations in superconducting condensation energy can be larger than its mean value.\cite{Zhou}  The physical origin of these exceedingly large fluctuations lies in the level statistics, precisely the origin of the UCF in normal metals.

The fluctuation in condensation energy will in turn manifest itself in fluctuations of the local $T_c$ even for a homogeneously disordered superconductor.\cite{Z-B}  Zhou and Biagini~\cite{Z-B} have shown that mesoscopic fluctuations of both Aslamasov-Larkin and Maki-Thompson contributions to conductivity would lead to a sample-specific conductance fluctuation above $T_c$.  Because of the long-range phase coherence developing in superconductors as $T_c$ is approached, sample-specific conductance should be observable in arbitrarily large samples, as long as the temperature is sufficiently close to $T_c$.  Similar to normal samples, these fluctuations are sensitive to magnetic field, impurity configuration, and gate voltage.  Conductance fluctuations are greatly amplified due to the superconducting coherence resulted from Cooper pairing correlation, a spectacular example of quantum mesoscopic phenomena at a macroscopic scale.

Experimental observations that might be related to those discussed here have been reported previously.  In particular, giant conductance fluctuations and MFPs have been reported for ultrashort quench-condensed Sn films.\cite{Dynes}  Negative magnetoresistance has also been observed in quench-condensed Pb wires.\cite{Xiong}.  Spivak and Kivelson~\cite{S-K} have shown earlier that negative MR can be naturally accounted for in the same theoretical framework as discussed above.  In addition, the observed asymmetry in MR may have a related physical origin.  The Spivak-Kivelson theory allows for time-reversal symmetry breaking in the ground state of a disordered superconductor, which can lead to asymmetric MR.

In conclusion, we have observed reproducible, sample-specific resistance fluctuations in disordered Au$_{0.7}$In$_{0.3}$ cylinders.  The amplitude of the fluctuation is much larger than that of the UCF in normal samples.  We have argued that the physical origin of these observations lies in the mesoscopic fluctuation of superconducting condensation energy, as predicted by theory.

The authors would like to acknowledge useful discussions with S.~Kivelson, B.~Spivak, and F.~Zhou.  This work is supported by NSF under grant DMR-9702661.


\section*{References}
\begin{thebibliography}{99}

\bibitem{1} For a review, see {\it Mesoscopic Phenomena in Solids}, eds. B.~L.~Altshuler, P.~A.~Lee, and R.~A.~Webb, (North-Holland, 1991).

\bibitem{2} C.~J.~Lambert and R.~Raimondi, {\em J. Phys.: Condens. Matter} {\bf
10}, 901 (1998), and references cited therein.

\bibitem{3} F.~Zhou, B.~Spivak, and A.~Zyuzin, \PRB {\bf 52}, 4467 (1995).

\bibitem{4} K.~Hecker {\it et al.}, \PRL {\bf 79}, 1547 (1997).

\bibitem{5} T.~T.~Heikkil\"a, M.~M.~Salomaa, and C.~J.~Lambert, \PRB {\bf 60}, 9291 (1999).

\bibitem{alloy} {\it Phase Diagrams of Binary Gold Alloys}, eds. H.~Okamoto and 
T.~B.~Massalski, (ASM International, 1987), pp. 142-153.

\bibitem{films} Yu.~Zadorozhny and Y.~Liu, to be published 
%preprint, cond-mat/9908281, (2000).

\bibitem{finger} Yu.~Zadorozhny, D.~R.~Herman, and Y.~Liu, preprint, cond-mat/9908281, (1999).

\bibitem{L-P} W.~A.~Little and R.~D.~Parks, \PRL {\bf 9}, 9 (1962).

\bibitem{Adams} W.~Wu and P.~Adams, \PRL {\bf 74}, 610 (1995).

\bibitem{6} See, for example, C.~P.~Umbach {\it et al.}, \PRB {\bf 30}, 4048 (1984).

\bibitem{Zhou} F.~Zhou, {\em Int. J. Mod. Phys.} {\bf 13}, 2229 (1999).

\bibitem{Z-B} F.~Zhou and C.~Biagini, \PRL {\bf 81}, 4724 (1998).

\bibitem{Dynes} A.~Frydman, E.~P.~Price, and R.~C.~Dynes, {\em Solid State Commun.} {\bf 106}, 715 (1998).

\bibitem{Xiong} P.~Xiong, A.~V.~Herzog, and R.~C.~Dynes, \PRL {\bf 78}, 927 (1997).

\bibitem{S-K} B.~Z.~Spivak and S.~A.~Kivelson, \PRB {\bf 43}, 3740 (1991); {\it ibid.} {\bf 45}, 10490 (1992).

\end{thebibliography}
\end{document}